\documentclass[conference]{IEEEtran}
\usepackage{amsmath,amsfonts,amsthm}
\usepackage[tikz,xcolor]{mdframed}
\usepackage{amsthm}

\newmdtheoremenv[
  roundcorner=5pt,
  linewidth=0.5pt,
  innerleftmargin=4pt,
  innerrightmargin=4pt,
  innertopmargin=0pt,
  innerbottommargin=4pt,
  skipabove=4pt,
  skipbelow=4pt
]{problem}{Problem}

\newtheorem{finding}{Finding}

\usepackage{algorithmic}

\usepackage[capitalize]{cleveref}
\crefname{table}{Tab.}{Tabs.}
\Crefname{problem}{Problem}{Problems}
\Crefname{finding}{Finding}{Findings}
\Crefname{table}{Table}{Tables}
\crefname{figure}{Fig.}{Figs.}
\Crefname{figure}{Figure}{Figures}
\crefname{section}{Sec.}{Secs.}
\Crefname{section}{Section}{Sections}
\crefname{algorithm}{Alg.}{Algs.}
\Crefname{algorithm}{Algorithm}{Algorithms}

\usepackage{booktabs}
\usepackage{multirow}
\usepackage{threeparttable}

\usepackage{graphicx}
\usepackage{subcaption}

\usepackage{tabularx}
\newcolumntype{Y}{>{\centering\arraybackslash}X}

\usepackage[dvipsnames]{xcolor}

\usepackage{colortbl}
\definecolor{LightGray}{gray}{0.8}

\usepackage{calc}
\newcommand{\textover}[3][l]{%
 \makebox[\widthof{#3}][#1]{#2}%
}

\usepackage{enumitem}

\usepackage{lipsum}

\usepackage{dirtree}

\usepackage[hang,flushmargin]{footmisc}
\renewcommand{\figurename}{\small Fig.}
\renewcommand{\tablename}{\small Tab.}
\usepackage{eso-pic}

\AddToShipoutPictureFG*{%
  \AtPageLowerLeft{%
    \raisebox{0.18in}{%
      \makebox[\paperwidth][c]{%
        \parbox{\textwidth}{%
          \centering
          \fontsize{6.3}{6.3}\selectfont
          \IEEEAcceptedCopyrightNotice
        }%
      }%
    }%
  }%
}

\newcommand{\IEEEAcceptedCopyrightNotice}{%
  © 2026 IEEE.  Personal use of this material is permitted.  Permission from IEEE must be obtained for all other uses, in any current or future media, including reprinting/republishing this material for advertising or promotional purposes, creating new collective works, for resale or redistribution to servers or lists, or reuse of any copyrighted component of this work in other works.
}
\usepackage{float}
\usepackage{cite}
\def\BibTeX{{\rm B\kern-.05em{\sc i\kern-.025em b}\kern-.08em
    T\kern-.1667em\lower.7ex\hbox{E}\kern-.125emX}}
\begin{document}

\title{
Evaluating Out-of-Distribution Robustness in Graph-Based Android Malware Classification:\\
A New Principled Benchmark
}

\author{
\IEEEauthorblockN{
Ngoc N. Tran\textsuperscript{1},
Anwar Said\textsuperscript{2},
Waseem Abbas\textsuperscript{3},
Tyler Derr\textsuperscript{1},
Xenofon D. Koutsoukos\textsuperscript{1}
}
\IEEEauthorblockA{
\textsuperscript{1}Vanderbilt University, USA \;\;\;\;
\textsuperscript{2}Highmark Health, USA \;\;\;\;
\textsuperscript{3}University of Texas at Dallas, USA \\
\{ngoc.n.tran,tyler.derr,xenofon.koutsoukos\}@vanderbilt.edu,
anwar.said@highmarkhealth.org, waseem.abbas@utdallas.edu
}
\vspace{-0.5cm}
}

\maketitle

\begin{abstract}
While graph-based Android malware classifiers report strong benchmark accuracy of over 94\%, their performance sharply decreases up to 45\% when exposed to previously unseen variants of known malware families. In this work, we systematically investigate this critical yet overlooked challenge for real-world deployment by introducing a benchmarking suite designed to simulate two prevalent scenarios: MalNet-Tiny-Common for covariate shift, and MalNet-Tiny-Distinct for domain shift.
We further identify an inherent limitation of existing benchmarks where input representation is limited to structure-only function call graphs, discarding the semantic signals needed for robust cross-distribution reasoning.
To verify this, we propose a semantic enrichment framework that extends raw graph topology with function-level attributes, combining lightweight metadata with LLM-based code embeddings.
Empirical evaluations confirm the effectiveness of our data-centric methodology, with which classification performs better under distribution shift compared to model-based approaches, and consistently further enhances robustness when used in conjunction.
We release our precomputed datasets alongside an extensible pipeline implementation, laying the groundwork for more resilient malware detection systems in evolving threat environments.
\end{abstract}


\vspace{-0.5em}
\section{Introduction}
\label{sec:introduction}

Android malware continues to evolve rapidly, posing persistent challenges to the reliability and robustness of automated detection systems. Recent advances in graph-based learning have led to promising approaches for malware classification by representing applications as \textit{function call graphs} (FCGs), where nodes correspond to individual functions and directed edges represent invocation relationships~\cite{bodmas,harang2020sorel20mlargescalebenchmark,joyce2021motiflargemalwarereference}. Graph Neural Networks (GNNs)~\cite{kipf2017semisupervisedclassificationgraphconvolutional, hamilton2018inductiverepresentationlearninglarge} applied to these graphs enable the modeling of structural and behavioral patterns that are indicative of malicious behavior~\cite{dmalnet}. Compared to image-based approaches like byteplot representations~\cite{freitas2022malnetlargescaleimagedatabase}, graph-based methods provide a more interpretable, semantically structured, and functionally meaningful abstraction.

Graph-based malware classification gained momentum with the release of MalNet~\cite{freitas2021largescaledatabasegraphrepresentation}, a large-scale dataset of over 1.5 million Android malware samples represented as FCGs. A smaller, balanced subset named MalNet-Tiny was released to support benchmarking. Since then, MalNet-Tiny has become a standard for evaluating GNN architectures, with state-of-the-art models achieving over 94\% accuracy~\cite{rampášek2023recipegeneralpowerfulscalable, shirzad2023exphormersparsetransformersgraphs}. However, these results often assume standard data splits. Recent work~\cite{wu2024graph} shows that performance drops significantly when evaluating models on samples from malware families not seen during training, highlighting a critical weakness under distribution shift. As prior research mostly focuses on achieving the highest accuracy on a well-conditioned dataset, this gap has been largely overlooked in the literature, severely limiting the practical deployment of graph-based classifiers in real-world scenarios where unseen malware variants frequently emerge.

To investigate, we construct and release two new benchmark variants of MalNet-Tiny for evaluating robustness under distribution shifts. In \emph{MalNet-Tiny-Common}, training and test samples are drawn from overlapping malware families but different subtypes, simulating \textit{covariance shift} where novel approaches to an existing attack paradigm are encountered. For \emph{MalNet-Tiny-Distinct}, test samples originate from families unseen during training, replicating \textit{domain shift} when completely unknown malware families often come up in practice. We thoroughly benchmark several state-of-the-art GNN architectures on these datasets, and evaluate the performance under distribution shift of existing solutions to domain generalization, to present the current development landscape.

Another key limitation underlying this sensitivity to distribution shifts is the lack of semantic information in MalNet-Tiny. During graph construction, all function-level data (e.g., names, types, code) was removed to avoid exposing potentially sensitive artifacts~\cite{freitas2021largescaledatabasegraphrepresentation}. While this decision aimed to reduce reverse-engineering risks, we argue it unnecessarily restricts the model's ability to reason about functional behavior. First, the full malware binaries used to generate these graphs are publicly available~\cite{androzoo, freitas2022malnetlargescaleimagedatabase}. Second, prior work has shown that semantic features can be integrated without compromising security~\cite{anderson2018emberopendatasettraining, bodmas}.
Moreover, we hypothesize that this omission impairs generalization for FCGs, whose edges denote information flow. 
In such graphs, node semantics are essential for effective message passing and representation learning~\cite{hamilton2018inductiverepresentationlearninglarge, hu2020ogb}. Without them, models tend to memorize local structures that do not transfer well across malware families.

Motivated by this insight, our second contribution proposes an enhanced attributed graph construction framework tailored for Android malware classification under distribution shift. Our method enriches each FCG with semantic node features extracted directly from the malware's code. Specifically, we extract a set of lightweight, widely available metadata features---including function names, signatures, access flags, instruction statistics, and Android-specific behaviors commonly used in malware analysis~\cite{anderson2018emberopendatasettraining, bodmas}. When decompiled source code is available, we embed function bodies using a large language model (LLM), providing dense representations of behavioral semantics. These semantic vectors are combined with standard structural features and injected as node attributes for downstream GNNs. Importantly, our framework is designed for real-world deployment scenarios, where semantic features may be partially missing. To address this, we introduce three collation strategies---\textit{Trim}, \textit{Zero}, and \textit{Prune}---that transform partially defined graphs into a consistent format suitable for learning.
We find that our proposed semantic graph construction significantly improves classification accuracy under both settings. These results highlight the critical role of semantic information in improving the generalization of graph-based malware classifiers under novel threats.

In summary, our key contributions are as follows:
\begin{itemize}
    \item We identify and demonstrate the brittleness of state-of-the-art graph-based Android malware classifiers under distribution shift, where detection accuracy drops sharply on test samples drawn from unseen malware families.

    \item We construct and release two new benchmark datasets, \emph{MalNet-Tiny-Common} and \emph{MalNet-Tiny-Distinct}, designed to evaluate classifier robustness under different types of distribution shifts.
    This includes a a semantic feature enrichment framework for Android FCGs that augments structural graphs with function metadata and LLM-derived embeddings, along with three collation strategies to mitigate real-world data quality challenges. 

    \item We evaluate the performance of different GNN architectures and training strategies on the new datasets with/without our proposed semantic graph representations, empirically demonstrating the effectiveness of semantic features against distribution shift when combined with adaptation-based training.
\end{itemize}

\noindent\textbf{Differences from existing datasets.}
There are very limited number of datasets for Android malwares available on the internet, especially when some cannot be found~\cite{8752028} or explicitly taken down~\cite{zhou2012dissecting}.
Most existing datasets extract features on a sample level~\cite{arp2014drebin,borah2020malware,mahdavifar2022effective,bragancca2025mh}, and/or lacked reproducible process for future feature engineering.
Others~\cite{makkawy2025malvislargescaleimagebasedframework,freitas2022malnetlargescaleimagedatabase} represent malwares as byteplots, effectively posing malware classification as an image classification task, and thus do not leverage the inherent structure of the input.
Closest to our work, \cite{GUERRAMANZANARES2021102399,haque2025lamdalongitudinalandroidmalware} have contributed datasets on \textit{concept drift}, where the definition of malwares changes over time; however, this direction is orthogonal to our versions of distribution shift. Moreover, these datasets also only extracted features on the sample level; and with unevenly distributed classes, they create additional challenges for continuing experimentation.

\section{Preliminaries}
\label{sec:preliminary}

\subsection{MalNet and Graph-Based Malware Data}

MalNet~\cite{freitas2021largescaledatabasegraphrepresentation} is a large-scale dataset of Android malware samples, where each sample is represented as a FCG extracted using AndroGuard~\cite{desnos2018androguard}. Each node in the FCG corresponds to a function, and edges indicate function calls. Labels are derived from VirusTotal reports and unified into a hierarchy of malware \emph{families} and \emph{types} using Euphony~\cite{euphony}. A cleaned and balanced subset, MalNet-Tiny, was introduced to support manageable training and evaluation, with each class defined by a unique (family, type) pair.

\subsection{Distribution Shift in Malware Classification}

We consider a standard supervised classification setting, where each malware sample is denoted by $X$, and its associated label by $Y$. The training data is drawn from a source distribution $\mathcal{D}$, while test-time samples may be drawn from a different target distribution $\mathcal{D}'$.
\textit{Distribution shift} then refers to the condition where the distributions of training and testing data differ, leading to a performance drop during model evaluation. While there are many types of distribution shifts~\cite{quinonero2022dataset}, we focus only on settings where labels for malwares do not change, i.e. $\mathbb{P}_{(X,Y)\sim\mathcal{D}}(Y|X)=\mathbb{P}_{(X,Y)\sim\mathcal{D}^\prime}(Y|X)$.
Among them, \textit{covariate shift} happens when malwares of the same labels are collected from different sources, meaning $\mathbb{P}_{\mathcal{D}}(Y)=\mathbb{P}_{\mathcal{D}^\prime}(Y)$ but $\mathbb{P}_{\mathcal{D}}(X)\ne\mathbb{P}_{\mathcal{D}^\prime}(X)$.
\textit{Domain shift} is a more general form of distribution shift where the training and test data come from completely different distributions, i.e. $\mathbb{P}_{\mathcal{D}}(Y)\ne\mathbb{P}_{\mathcal{D}^\prime}(Y)$ and $\mathbb{P}_{\mathcal{D}}(X)\ne\mathbb{P}_{\mathcal{D}^\prime}(X)$.
\textit{Temporal shift} is similar to covariate shift, but differs in that the distribution mismatch is caused by malwares changing over time.

\subsection{Research Objectives}
Malware classifiers must remain effective under distribution shift, as new variants often differ from prior data. 
Thus, with the goal of benchmarking this problem, we introduce two MalNet-Tiny-style datasets simulating covariate and domain shifts, and propose a framework that enhances attributed graph construction and mitigates data-quality issues.

\section{Constructing Distribution-Shifted Datasets}\label{sec:datasets}
\begin{figure}[t]
\centering
\includegraphics[width=\linewidth]{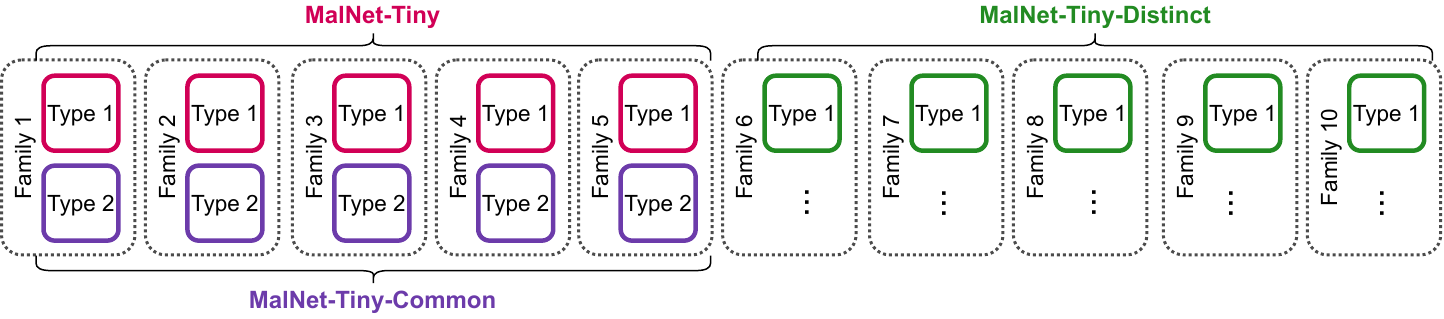}
\caption{Diagram of the dataset labels. Dashed boxes (columns) denote malware families, solid box denotes types/classes.}
\label{fig:classes}
\vspace{-1.5ex}
\end{figure}

In this section, we describe the construction of our new benchmark datasets.
Recall that each malware label in MalNet-Tiny is a malware \textit{family/type} pair, where \textit{family} is a broader category~of malwares with similar characteristics, and \textit{type} is a subcategory within a family. Leveraging this hierarchical labeling, we construct new datasets from the original {MalNet} to realistically simulate different distribution shifts from MalNet-Tiny:
MalNet-Tiny-Common for covariate shift,
and MalNet-Tiny-Distinct for domain shift.
These datasets are curated to have these same properties for both compatibility and a fair comparison with the baselines.
Malware families/types and their corresponding samples are selected to be as disjoint as possible, with the chosen labels listed in \cref{tab:malware_types}. \cref{fig:classes} visualizes the relationships between the labels across datasets.
We also intended to create a dataset for temporal shift, but were unable to do so due to MalNet having too few malwares belonging to MalNet-Tiny classes despite its large size. While we believe our proposed method is also robust to temporal shift, we leave this as a future work.

\begin{figure*}[t]
\centering
\includegraphics[width=\linewidth]{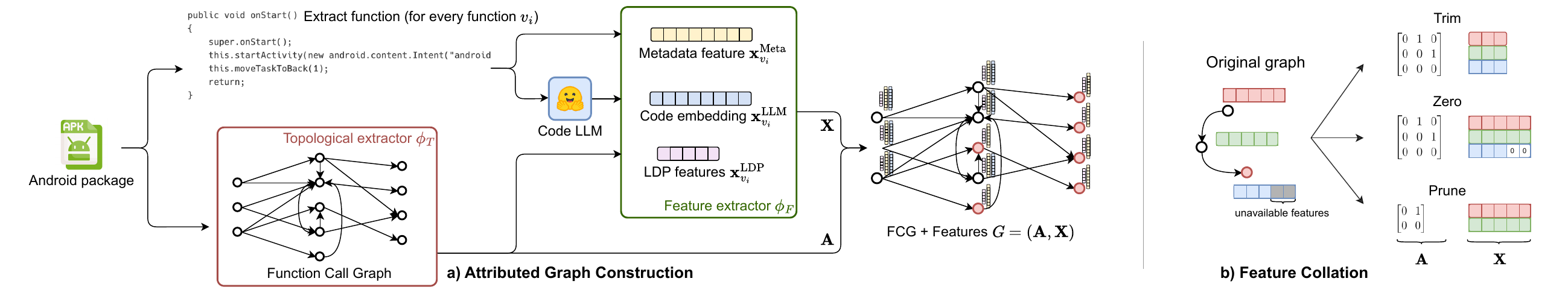}
\caption{Overview of the robust graph learning pipeline. For each Android package, we \textit{(a)} extract and merge the FCG structure and semantic function features into a complete attributed graph, where red nodes indicate unavailable features (\cref{sec:feature-extraction}); then \textit{(b)} make sure all input graphs with missing features are model-ready (\cref{sec:feature_collation}).}
\label{fig:method}
\vspace{-1ex}
\end{figure*}

\begin{table}[ht]
\small
\centering
\caption{Malware labels of MalNet-Tiny variants (\textit{family/type}).}
\begin{tabularx}{\columnwidth}{lll}
\toprule
\textcolor{RubineRed}{\textbf{MalNet-Tiny}} & \textcolor{RoyalPurple}{\textbf{MNT-Common}} & \textcolor{ForestGreen}{\textbf{MNT-Distinct}} \\
\midrule
{addisplay}{} / kuguo & {addisplay}{} / dowgin & {spr}{} / lootor \\
{adware}{} / airpush & {adware}{} / startapp & clicker+trj / dowgin \\
{benign}{} / benign & {benign}{} / benign & {riskware}{} / nandrobox \\
downloader / jiagu & downloader / \textit{mixed} & {malware}{} / \textit{mixed} \\
{trojan}{} / artemis & {trojan}{} / deng & {spyware}{} / \textit{mixed} \\
\bottomrule
\end{tabularx}
\label{tab:malware_types}
\vspace{-1ex}
\end{table}

\subsection{Covariate Shift: MalNet-Tiny-Common}
We construct MalNet-Tiny-Common to have the same malware families but different malware types to MalNet-Tiny to simulate a covariate shift scenario. The label correlation between the two datasets let us alternatively interpret the training process as fitting malwares to their corresponding family labels instead of types, implying that a model trained on MalNet-Tiny can be used as-is to classify MalNet-Tiny-Common, but performance will drop due to the distribution shift.
For the sampling process, seeding randomness with 0, we start by sampling types from the same families as MalNet-Tiny, and then sample 1,000 samples from each type. As the benign family doesn't have multiple types; and thus the new benign split
is simply sampled to not have any overlaps.

\subsection{Domain Shift: MalNet-Tiny-Distinct}
MalNet-Tiny-Distinct is composed similar to MalNet-Tiny-Common, but with different malware families/types to simulate domain shift. This dataset is designed to test the adaptation schemes' ability to generalize to completely novel malwares, and thus is more challenging than MalNet-Tiny-Common. As the Android repository did not have enough malware families with pure single-type despite its large size, we select our class splits to have minimum type entropy per family.

\section{Feature Construction and Data Processing}
\label{sec:method}

As hypothesized in \cref{sec:introduction}, we believe that the lack of semantic information in MalNet-Tiny significantly contributes to the brittleness of graph-based malware classifiers under distribution shift. To address this, we propose a comprehensive pipeline for robust malware classification to train a graph classification model for malware samples, as illustrated in \cref{fig:method}.
We first reformulate the problem of malware classification as a graph classification task, and present our proposed method for constructing attributed graphs from malware samples. We then discuss data-quality challenges in the original MalNet that hinders the feature extraction process, and how we can mitigate them with our proposed collation schemes.

\subsection{Formalizing Robust Feature Extraction}
We formally define malware classification as a graph classification task:
For each malware sample $X$, we aim to construct an attributed graph
$G = (\mathbf{A}, \mathbf{X})$, where $\mathbf{A}=\phi_T(X)$ is the adjacency matrix, and $\mathbf{X}=\phi_F(X)$ is the node feature matrix
\footnote{For each malware sample, normal $X$ refers to the raw sample, $G$ refers to its graph representation, and bold $\mathbf{X}$ refers to the node feature matrix.}.
This process involves a topology extractor $\phi_T$ that extracts the graph structure from the malware sample, and a node feature extractor $\phi_F$ that extracts a feature vector for each vertex in the graph; which we combined as $\phi$.
The goal is to construct a semantic feature extractor $\phi$ such that the trained graph classification model $\hat{F}$ performs well on both the training distribution $\mathcal{D}$ and a new testing distribution $\mathcal{D}^\prime$:

\begin{problem}[Robust Feature Extraction for Graph Malware Classification]
Given malware sample $X$, the goal is to construct an expressive feature extractor $\phi(X)$ that returns the graph representations $G=\phi(X)$, such that the trained model $\hat{F}$ performs well on both the training,
i.e. high $\mathbb{P}_{(X,Y)\sim\mathcal{D}}[\hat{F}(G)=Y]$,
and new test distribution,
i.e. high\linebreak $\mathbb{P}_{(X,Y)\sim\mathcal{D}^\prime}[\hat{F}(G)=Y]$.
\end{problem}

\vspace{-1ex}
\subsection{Attributed Graph Construction}
\label{sec:feature-extraction}

In this section, we present our proposed method for constructing attributed graphs from malware samples focused on advanced feature extraction that we believe may aid in mitigating the distribution shift problem.
\cref{fig:method} shows an overview of the feature construction process.

\subsubsection{Topology Extractor}
We use FCGs for the topology following MalNet, which contains
a node set $\mathcal{V}$ where each node $v_i$ represents a function, and an edge set $\mathcal{E}$ where each edge $e_{jk}$ denotes that function $v_j$ calls function $v_k$ during its execution. Mathematically, we define the topology extractor $\phi_T$ as a function that takes in a malware sample $x$ and returns the adjacency matrix $\mathbf{A}=\phi_T(X)\in\mathbb{R}^{|\mathcal{V}|\times |\mathcal{V}|}$ of the FCG, where $|\mathcal{V}|$ is the number of nodes/functions, and each entry $A_{jk}$ is 1 if $e_{jk}\in \mathcal{E}$, and 0 otherwise. The resulting graph is then $G=(\mathbf{A}, \emptyset)$, where the node feature matrix is empty.

\subsubsection{Feature Extractor}
Existing methods~\cite{freitas2021largescaledatabasegraphrepresentation,rampášek2023recipegeneralpowerfulscalable,shirzad2023exphormersparsetransformersgraphs} typically use Local Degree Profile (LDP)~\cite{cai2022simpleeffectivebaselinenonattributed} for node feature extraction $\phi_F$, which is a vector of the node's degree and its neighbors' degree statistics. 
As LDP focuses solely on the structural properties of the graph,
it unnecessarily limits the model's ability to learn from the semantic properties of these function calls.
Instead, we propose to extract node features directly from the corresponding functions, which better capture the behavior and characteristics of the malwares.

\noindent\textbf{\textit{Aggregated Metadata Features.}}
Inspired by EMBER~\cite{anderson2018emberopendatasettraining}, we construct a set of metadata features for each function, which are then aggregated into a single feature vector $\mathbf{x}^\text{Meta}_{v_i}$ for each node $v_i$. Features include \emph{class} and \emph{method names, method signatures, access flags, code length, bytecode statistics, instruction statistics, string statistics, and more}. In addition, we extract other Android-specific features such as \emph{storage access, registry modifications, and in-memory code execution}.
This feature construction scheme is designed to capture the essential characteristics of the functions in the FCGs, while not disclosing any sensitive information similar to the EMBER precedence.
We refer the reader to Appendix~\ref{sec:metafeats} for details.

\noindent\textbf{\textit{LLM Source Code Embedding.}} 
\label{sec:llm}
From the Androguard analysis output, we obtain the decompiled Java source, which we use to extract additional features. We utilize CodeXEmbed~\cite{liu2024codexembedgeneralistembeddingmodel}
(the only available code embedding model with large enough context window) to embed the source code into a numerical vector. 
The model~\cite{wolf2020huggingfacestransformersstateoftheartnatural} is pulled from HuggingFace and inference was done through their provided API.
This embedding better extracts the semantic meaning of the function than the aggregated metadata features alone, as it captures the function's behavior and purpose based on its implementation. We denote the resulting feature vector as $\mathbf{x}^\text{LLM}_{v_i}$ for each function $v_i$. Further details and additional experiments on different LLMs are included in Appendix~\ref{sec:llm_ablation}.

\noindent\textbf{\textit{Concatenating Components.}}
While the metadata contains some general properties of a function,
incorporating code embeddings generated by LLMs provides 
more comprehensive function behavior.
This provides additional context for the model to distinguish between functions with similar metadata.
This insight leads us to include both types of features in our method. We also utilize the LDP features, which were originally employed in previous works~\cite{shirzad2023exphormersparsetransformersgraphs,freitas2021largescaledatabasegraphrepresentation,rampášek2023recipegeneralpowerfulscalable}:

\vspace{-3ex}
\begin{align}
\mathbf{x}_{v_i}^\text{LDP}
= [\mathrm{deg}(v_i), \min_{v_j\in \mathcal{N}(v_i)}\mathrm{deg}(v_j), \max_{v_j\in \mathcal{N}(v_i)}\mathrm{deg}(v_j), \\ \nonumber
\underset{v_j\in \mathcal{N}(v_i)}{\mathrm{mean}}\mathrm{deg}(v_j), \underset{v_j\in \mathcal{N}(v_i)}{\mathrm{std}}\mathrm{deg}(v_j)],
\end{align}

\vspace{-1ex}
\noindent for any node $v_i$, where $\mathrm{deg}(v_i)$ is the node degree, and $\mathcal{N}(v_i)$ is the neighborhood of $v_i$.

The three extracted feature vectors are concatenated to form the final $d$-dimensional node feature vector:

$$\mathbf{x}_{v_i} = \begin{bmatrix}\mathbf{x}_{v_i}^\text{Meta} & \mathbf{x}_{v_i}^\mathrm{LLM} & \mathbf{x}_{v_i}^\text{LDP}\end{bmatrix}\in\mathbb{R}^d,$$ 
for each node $v_i$ in the FCG. Putting it all together, we define the feature extractor $\phi_F$ as a function that takes a malware $X$ and its the FCG adjacency matrix $\mathbf{A}$, and returns the node feature matrix
$\mathbf{X}=\phi_F(\mathbf{A}, X)=[\mathbf{x}_{v_1}\;\dots\;\mathbf{x}_{v_{|\mathcal{V}|}}]^\top\in\mathbb{R}^{|\mathcal{V}|\times d}$, with each row vector $\mathbf{x}_{v_i}$ constructed as described above. As $\mathbf{A}=\phi_T(X)$ can be obtained from the malware sample $X$ and thus can be omitted from the function signature, we henceforth omit it for simplicity.
The final attributed graph is then:

\vspace{-1.5ex}
\begin{equation}
G=\phi(X)=(\phi_T(X), \phi_F(X))=(\mathbf{A}, \mathbf{X}).
\end{equation}

\vspace{-1ex}
\subsection{Handling Data-Quality Challenges}
\label{sec:feature_collation}
While the topology extractor $\phi_T$ produces a well-defined graph structure $\mathbf{A}$ for each malware sample, constructing a high-quality node feature matrix $\mathbf{X} = \phi_F(X)$ presents nontrivial data-quality challenges. In particular, the reliability and availability of semantic features extracted from each function in the graph can vary substantially due to limitations in the underlying data.
For any given malware sample, the number of extractable semantic features varies from function to function. Some nodes correspond to external/system functions (e.g., Android APIs or libraries) that lack accessible decompiled source code, making it impossible to compute certain semantic features such as code embeddings. As a result, the feature matrix $\mathbf{X} \in \mathbb{R}^{|\mathcal{V}| \times d}$, where each row $\mathbf{X}_{i,:}\equiv \mathbf{x}_{v_i} \in \mathbb{R}^d$ represents the feature vector for node $v_i$, is only partially defined, with some node feature entries missing due to limitations in static analysis or obfuscation.
To formalize this, we define the set of \textit{non-universal} feature dimensions:

\vspace{-3.5ex}
\begin{align*}
\mathcal{C} = \left\{ c \in \{1, \dots, d\} \,\middle|\, \exists\, i \in \{1, \dots, |\mathcal{V}|\} \text{ s.t. } \mathbf{X}_{i,c} \text{ is missing} \right\}.
\end{align*}

\vspace{-1ex}
\noindent These are the feature types that are not consistently available across all nodes in this sample's graph.
However, since the downstream GNN classifier $F(G)$ requires that all nodes in the attributed graph $G = (\mathbf{A}, \mathbf{X})$ share a consistent input dimension, we must address this structural inconsistency before training. To mitigate this challenge, we propose three node feature collation schemes designed to transform the partially defined $\mathbf{X}$ into a complete, uniform format compatible with GNN-based learning: \textit{Trim}, \textit{Zero}, and \textit{Prune}.

\subsubsection{\textit{Trim}: Remove Non-Universal Feature Dimensions}
We remove all feature dimensions that are not available for all nodes in the current sample. That is, we retain only the set of feature dimensions $\mathcal{F}$ that are defined across every node, i.e., $\mathcal{F} = \{1, \dots, d\} \setminus \mathcal{C}.$ We then construct the trimmed feature matrix as follows:

\vspace{-3.5ex}
\begin{align*}
\mathbf{X}_{\text{trim}} = \mathbf{X}_{:, \mathcal{F}} \in \mathbb{R}^{|\mathcal{V}| \times |\mathcal{F}|}.
\end{align*}

\vspace{-1ex}
\noindent This ensures every node has complete/consistent features. However, it discards any partially defined features, potentially removing useful information available for some functions.

\subsubsection{\textit{Zero}: Impute Missing Features with Zeros}
We retain the full graph and all feature dimensions, and fill in any missing values with zeros. Let:

\vspace{-3ex}
\begin{align}
\mathcal{M} = \left\{ (i, c) \in \{1, \dots, |\mathcal{V}|\} \times \mathcal{C} \,\middle|\, \mathbf{X}_{i,c} \text{ is missing} \right\} \nonumber
\end{align}

\vspace{-0.75ex}
\noindent be the index set of node-feature pairs that are undefined. We define the zero-imputed feature matrix $\mathbf{X}_{\text{zero}} \in \mathbb{R}^{|\mathcal{V}| \times d}$ as:

\vspace{-3ex}
\begin{align}
\mathbf{X}_{\text{zero}}[i,c] =
\begin{cases}
\mathbf{X}[i,c] & \text{if } (i,c) \notin \mathcal{M}, \\
0 & \text{if } (i,c) \in \mathcal{M}.
\end{cases}\nonumber
\end{align}

\vspace{-0.75ex}
\noindent This approach ensures that every node retains a full-length feature vector, enabling direct compatibility with GNN input requirements. It defers to the model to learn whether zeroed features are informative or irrelevant.

\subsubsection{\textit{Prune}: Discard Nodes with Incomplete Feature Vectors}
Unlike other strategies which retain the original graph topology, \textit{Prune} modifies both the feature matrix and the graph structure: when a node has missing feature values, we remove it along with any edges connected to it.
Formally, we first define the set of nodes with complete feature vectors $\mathcal{V}^\prime$:

\vspace{-3ex}
\begin{align}
\mathcal{V}^\prime = \left\{ v_i \in \mathcal{V} \,\middle|\, \forall c \in \{1, \dots, d\},\ \mathbf{X}_{i,c} \text{ is defined} \right\}. \nonumber
\end{align}

\vspace{-0.75ex}
\noindent We then restrict both the feature matrix and the adjacency matrix to this subset. Our newly pruned feature matrix is: 

\vspace{-3.5ex}
\begin{align}
\mathbf{X}_{\text{prune}} = \mathbf{X}_{\mathcal{V}^\prime, :} \in \mathbb{R}^{|\mathcal{V}^\prime| \times d}, \nonumber
\end{align}

\vspace{-1ex}
\noindent and the corresponding pruned graph topology is:

\vspace{-3ex}
\begin{align}
\mathbf{A}_{\text{prune}} = \mathbf{A}_{\mathcal{V}^\prime, \mathcal{V}^\prime} \in \{0,1\}^{|\mathcal{V}^\prime| \times |\mathcal{V}^\prime|}. \nonumber
\end{align}

\vspace{-1ex}
\noindent This operation yields the subgraph of $G $ induced by the node set $( \mathcal{V}^\prime )$, thereby ensuring that the input graph provided to the GNN maintains structural consistency and contains fully-defined features.
While preserving all feature dimensions, it might omit important contextual information.

\section{Graph Learning for Distribution Shifts}
Here we finally leverage GNNs to learn malware representations based on our enhanced attributed graphs:
with each malware transformed into an attributed graph $G = (\mathbf{A}, \mathbf{X})$, it proceeds to be encoded into a condensed graph-level vector representation via a GNN $F(G)$.
Pursuing the goal of building models that remain robust when the test-time data distribution differs from the training, i.e., \textit{distribution shift},
we further improve model generalizability by applying existing model-centric adaptation approaches jointly with our data-centric enrichment scheme.
To integrate these model-based techniques into our pipeline, we decompose the learning process into two stages: \textit{upstream training} on the source distribution $\mathcal{D}$, and \textit{downstream adaptation} to the target distribution $\mathcal{D}^\prime$.

\subsubsection{Upstream Training}
In the upstream phase, we assume access to labeled training samples $(G, Y)\sim\mathcal{D}$, where each malware has been converted into an attributed graph. We train a GNN classifier $F$ by minimizing the expected loss:

\vspace{-3ex}
\begin{align}
\hat{F} = \arg\min_F \, \mathbb{E}_{(G, Y) \sim \mathcal{D}} \left[ \mathcal{L}(F(G), Y) \right] \nonumber 
\end{align}

\vspace{-0.75ex}
\noindent This training step may involve supervised or self-supervised (pre-) training and does not assume any knowledge of $\mathcal{D}^\prime$.

\subsubsection{Downstream Adaptation}
At test time, we apply $\hat{F}$ to data drawn from a different distribution $\mathcal{D}^\prime$, which can result in degraded performance due to distribution shift. To improve generalization, we define a general adaptation framework that produces an adapted model $\tilde{F}$ by modifying $\hat{F}$ using data from $\mathcal{D}^\prime$, categorized by the target labels' availability.

\paragraph{Test-Time Adaptation (TTA).}
TTA assumes access to \textit{unlabeled} graphs from the target domain $\{G_j\}_{j=1}^m \sim \mathcal{D}^\prime$, and adapts the model using these samples:

\vspace{-3ex}
\begin{align}
\tilde{F} = \mathcal{A}_{\text{TTA}}(\hat{F}; \{G_j\}_{j=1}^m).
\end{align}

\vspace{-1ex}
\noindent These methods typically do not modify the entire model but update specific components such as normalization layers or classifier prototypes~\cite{t3a}. TTA is appealing in scenarios where no new annotations are available.

\paragraph{Domain Adaptation (DA).}
DA assumes access to \textit{labeled} target data $\{(G_k, Y_k)\}_{k=1}^n \sim \mathcal{D}^\prime$ and adapts the model using both graphs and their labels:

\vspace{-3ex}
\begin{align}
\tilde{F} = \mathcal{A}_{\text{DA}}(\hat{F}; \{(G_k, Y_k)\}_{k=1}^n).
\end{align}

\vspace{-1ex}
\noindent This setting supports stronger forms of adaptation, including full finetuning~\cite{church2021emerging}, parameter-efficient updates~\cite{gui2023gadapterstructureawareparameterefficienttransfer,han2024parameterefficientfinetuninglargemodels}, or classifier replacement~\cite{chen2020simpleframeworkcontrastivelearning}, but requires additional labeling.
\section{Numerical Evaluation}\label{sec:experiments}
In this section, we evaluate the performance of current graph-based malware classifiers in the presence of distribution shift on our newly constructed datasets, and investigate the effectiveness of our proposed semantic feature enrichment framework across different GNN architectures and adaptation methods.
For the latter goal,
we pose a series of research questions for a thorough evaluation:

\noindent\textbf{RQ1:} \textit{How do semantic features affect model robustness?}

We address this question by comparing models
trained with and without semantic features on the original dataset, evaluated on the same testing distribution (i.e. high utility), and on a covariate-shifted testing distribution (i.e. high robustness).

\noindent\textbf{RQ2:} \textit{Are all features necessary for improving robustness?}

We conduct an ablation study on the two components of our semantic feature construction with
different collation schemes.

\noindent\textbf{RQ3:} \textit{Do features work with existing adaptation methods?}

Given that our method takes a different approach to alleviating distribution shift, we evaluate how well it can be combined with and improve existing adaptation methods.

\subsection{Experimental Setup}
Depending on the model architecture, different feature collation schemes may be better or worse. Therefore, we conduct our experiments with various GNN architectures, including
GCN~\cite{kipf2017semisupervisedclassificationgraphconvolutional},
GIN~\cite{xu2019powerfulgraphneuralnetworks},
{GPS}~\cite{rampášek2023recipegeneralpowerfulscalable}, and
{Exphormer}~\cite{shirzad2023exphormersparsetransformersgraphs}.
Regarding adaptation methods, 
we select these methods such that they do not alter the upstream training process, and cover a wide range of approaches. For Test-time Adaptation (TTA), the selected baselines are {Tent}~\cite{wang2021tentfullytesttimeadaptation}, {T3A}~\cite{t3a}, and {GTrans}~\cite{jin2023empowering}.
For Domain Adaptation (DA) approaches, we evaluate with k-NN Probe~\cite{chen2020simpleframeworkcontrastivelearning}, normal finetuning, and {AdapterGNN}~\cite{li2023adaptergnnparameterefficientfinetuningimproves}. 

\newcommand{\greyno}{\textcolor{LightGray}{-}}
\begin{table*}[t]
 \small
\scriptsize
\centering
\caption{\small Models' accuracy on \textcolor{RubineRed}{MalNet-Tiny} and \textcolor{RoyalPurple}{MalNet-Tiny-Common} across different feature configurations. Top 5 highest values are highlighted in green, darker green represents higher accuracy. Subscript denotes standard deviation over 3 independent runs.}
\vspace{-0.5ex}
\renewcommand{\arraystretch}{0.7}
\begin{tabularx}{\textwidth}{c!{\color{LightGray}\vrule width 0.1pt}ccc!{\color{LightGray}\vrule width 0.1pt}YYYYYYYY}
\toprule
\multirow{2}{*}{Method} & \multicolumn{3}{c!{\color{LightGray}\vrule width 0.1pt}}{Features} & \multicolumn{2}{c}{GCN} & \multicolumn{2}{c}{GIN} & \multicolumn{2}{c}{GPS} & \multicolumn{2}{c}{Exphormer} \\
\cmidrule(lr){2-4} \cmidrule(lr){5-6} \cmidrule(lr){7-8} \cmidrule(lr){9-10} \cmidrule(lr){11-12}
 &  Meta  &  LLM  &  LDP  & \textcolor{RubineRed}{Tiny} & \textcolor{RoyalPurple}{Cmn.} & \textcolor{RubineRed}{Tiny} & \textcolor{RoyalPurple}{Cmn.} & \textcolor{RubineRed}{Tiny} & \textcolor{RoyalPurple}{Cmn.} & \textcolor{RubineRed}{Tiny} & \textcolor{RoyalPurple}{Cmn.} \\
\midrule

\multicolumn{3}{c!{\color{LightGray}\vrule width 0.1pt}}{Baseline / None} & \checkmark
 & $85.67_{0.58}$ & $48.13_{0.26}$ & $90.40_{1.44}$ & $47.43_{0.48}$ & $93.53_{0.09}$ & $48.23_{0.29}$ & $93.27_{0.49}$ & $48.40_{0.24}$\\
\midrule
\multirow{2}{*}{Trim}
& \checkmark & \greyno    & \greyno    &  $91.47_{0.39}$ &  $55.20_{0.99}$ &  $91.13_{0.66}$ &  $51.67_{0.53}$ &  $94.30_{0.16}$ &  $52.90_{0.78}$ &  $94.50_{0.45}$ &  \cellcolor{green!80}
$56.43_{2.24}$\\
& \checkmark & \greyno    & \checkmark &  $91.53_{0.40}$ &  $51.07_{0.45}$ &  $92.20_{0.24}$ &  $51.73_{1.00}$ &  $94.33_{0.69}$ &  \cellcolor{green!100}
$\mathbf{56.87_{2.03}}$ &  \cellcolor{green!100}
$\mathbf{95.00_{0.22}}$ &  $51.57_{1.32}$\\
\midrule
\multirow{7}{*}{Prune}
& \checkmark & \greyno    & \greyno    &  $93.33_{0.09}$ &  $54.40_{0.50}$ &  $92.73_{0.70}$ &  $55.17_{1.25}$ &  $93.63_{0.26}$ &  $54.47_{1.16}$ &  $94.40_{0.37}$ &  $51.73_{1.31}$\\
& \checkmark & \greyno    & \checkmark &  $92.87_{0.39}$ &  $51.30_{1.70}$ &  $93.47_{0.54}$ &  $54.60_{0.71}$ &  $94.27_{0.77}$ &  $54.37_{1.28}$ &  $94.30_{0.70}$ &  $54.70_{2.07}$\\
& \greyno    & \checkmark & \greyno    &  \cellcolor{green!20}
$94.33_{0.24}$ &  \cellcolor{green!100}
$\mathbf{60.30_{1.98}}$ &  $94.30_{0.43}$ &  \cellcolor{green!80}
$61.10_{0.22}$ &  \cellcolor{green!20}
$94.77_{0.26}$ &  $53.50_{2.12}$ &  $94.50_{0.22}$ &  $55.27_{0.66}$\\
& \greyno    & \checkmark & \checkmark &  \cellcolor{green!20}
$94.33_{0.29}$ &  $58.33_{1.72}$ &  \cellcolor{green!60}
$94.83_{0.42}$ &  \cellcolor{green!40}
$59.97_{0.95}$ &  \cellcolor{green!40}
$94.80_{0.28}$ &  $53.07_{0.73}$ &  \cellcolor{green!40}
$94.97_{0.52}$ &  $54.87_{1.52}$\\
& \checkmark & \checkmark & \greyno    &  \cellcolor{green!100}
$\mathbf{95.17_{0.33}}$ &  \cellcolor{green!80}
$59.70_{0.41}$ &  $94.20_{0.24}$ &  \cellcolor{green!100}
$\mathbf{61.63_{2.19}}$ &  \cellcolor{green!60}
$94.87_{0.41}$ &  \cellcolor{green!40}
$55.57_{1.52}$ &  \cellcolor{green!100}
$\mathbf{95.00_{0.43}}$ &  $52.37_{0.12}$\\
& \checkmark & \checkmark & \checkmark &  $94.23_{0.09}$ &  $57.03_{1.08}$ &  \cellcolor{green!40}
$94.70_{0.08}$ &  \cellcolor{green!60}
$60.07_{1.67}$ &  \cellcolor{green!20}
$94.77_{0.21}$ &  \cellcolor{green!60}
$56.70_{0.93}$ &  \cellcolor{green!40}
$94.97_{0.31}$ &  $53.93_{1.05}$\\
\midrule
\multirow{7}{*}{Zero}
& \checkmark & \greyno    & \greyno    &  $93.60_{0.37}$ &  $54.27_{0.79}$ &  $93.30_{0.73}$ &  $52.83_{0.74}$ &  $94.37_{0.25}$ &  $54.90_{1.41}$ &  $93.67_{0.29}$ &  $53.50_{2.94}$\\
& \checkmark & \greyno    & \checkmark &  $92.57_{0.26}$ &  $50.20_{0.99}$ &  $93.67_{0.34}$ &  $53.40_{1.28}$ &  $94.20_{0.36}$ &  $50.87_{0.87}$ &  $94.00_{0.51}$ &  \cellcolor{green!100}
$\mathbf{57.03_{0.46}}$\\
& \greyno    & \checkmark & \greyno    &  \cellcolor{green!80}
$94.60_{0.22}$ &  $56.70_{2.03}$ &  $94.27_{0.26}$ &  $58.60_{1.70}$ &  \cellcolor{green!80}
$95.10_{0.36}$ &  \cellcolor{green!80}
$56.73_{0.54}$ &  $94.60_{0.43}$ &  \cellcolor{green!40}
$55.57_{1.58}$\\
& \greyno    & \checkmark & \checkmark &  \cellcolor{green!60}
$94.57_{0.41}$ &  \cellcolor{green!20}
$58.60_{1.19}$ &  \cellcolor{green!20}
$94.50_{0.36}$ &  $56.57_{0.34}$ &  $94.60_{0.37}$ &  $54.87_{1.39}$ &  \cellcolor{green!100}
$\mathbf{95.00_{0.22}}$ &  \cellcolor{green!20}
$55.33_{2.33}$\\
& \checkmark & \checkmark & \greyno    &  $93.80_{0.16}$ &  \cellcolor{green!40}
$58.87_{1.46}$ &  \cellcolor{green!100}
$\mathbf{94.93_{0.34}}$ &  $56.33_{0.62}$ &  \cellcolor{green!100}
$\mathbf{95.27_{0.21}}$ &  \cellcolor{green!20}
$54.97_{1.48}$ &  \cellcolor{green!40}
$94.97_{0.47}$ &  \cellcolor{green!60}
$55.87_{0.31}$\\
& \checkmark & \checkmark & \checkmark &  \cellcolor{green!40}
$94.37_{0.37}$ &  \cellcolor{green!60}
$59.00_{0.86}$ &  \cellcolor{green!100}
$\mathbf{94.93_{0.29}}$ &  \cellcolor{green!20}
$59.10_{0.67}$ &  $94.53_{0.09}$ &  $53.60_{0.67}$ &  $94.40_{0.08}$ &  $54.73_{1.43}$\\
\bottomrule
\end{tabularx}
\label{tab:ablation}
\vspace{-2ex}
\end{table*}

\begin{table}[ht]
\scriptsize 
\centering
\caption{\small Test accuracy on \textcolor{RubineRed}{MalNet-Tiny} and \textcolor{RoyalPurple}{MalNet-Tiny-Common} with/without semantic features. Highest values are bolded.}
\vspace{-0.5ex}
\renewcommand{\arraystretch}{0.7}
\begin{tabularx}{\columnwidth}{cYYYYYYYY}
\toprule
\multirow{2}{*}{Method}& \multicolumn{2}{c}{GCN} & \multicolumn{2}{c}{GIN} & \multicolumn{2}{c}{GPS} & \multicolumn{2}{c}{Exphormer} \\
\cmidrule(lr){2-3} \cmidrule(lr){4-5} \cmidrule(lr){6-7} \cmidrule(lr){8-9}
& \textcolor{RubineRed}{Tiny} & \textcolor{RoyalPurple}{Cmn.} & \textcolor{RubineRed}{Tiny} & \textcolor{RoyalPurple}{Cmn.} & \textcolor{RubineRed}{Tiny} & \textcolor{RoyalPurple}{Cmn.} & \textcolor{RubineRed}{Tiny} & \textcolor{RoyalPurple}{Cmn.} \\
\midrule
Baseline & 85.7\% & 48.1\% & 90.4\% & 47.4\% & 93.5\% & 48.2\% & 93.3\% & 48.4\% \\
\midrule
Trim & 91.5\% & 55.2\% & 92.2\% & 51.7\% & 94.3\% & \textbf{56.9\%} & \textbf{95.0\%} & 56.4\% \\
Prune & \textbf{95.2\%} & \textbf{60.3\%} & 94.8\% & \textbf{61.6\%} & 94.9\% & 56.7\% & \textbf{95.0\%} & 55.3\% \\
Zero & 94.6\% & 59.0\% & \textbf{94.9\%} & 59.1\% & \textbf{95.3\%} & 56.7\% & \textbf{95.0\%} & \textbf{57.0\%} \\
\bottomrule
\end{tabularx}
\label{tab:rq1}
\vspace{-2ex}
\end{table}

\subsection{Results \& Findings}
\label{sec:results}

\subsubsection{Standard Training Under Covariate Shift}

\cref{tab:rq1} shows the results of our experiments on the original MalNet-Tiny and its covariate-shifted counterpart MalNet-Tiny-Common. With no semantic features, for all GNN architectures, accuracy drops by almost half when evaluated on the shifted distribution, up to a difference of 45.3\% in the case of GPS. This confirms the severity of distribution shift in encountering real-world malwares, and motivates the need for robust models that can generalize well to unseen data.
We next investigate how our semantic feature construction affects model robustness.

\noindent{\textbf{RQ1:} \textit{How do semantic features affect model robustness?}}

We observe that every models trained with semantic features outperform when compared to the baselines trained without.

\vspace{-0.5ex}
\begin{finding}
\textit{Prune} achieves the highest accuracy for traditional message-passing architectures (MPNNs), while \textit{Zero} achieves the highest accuracy for Transformer-based ones.
\end{finding}
\vspace{-0.5ex}

As \textit{Trim} achieves a lower accuracy for most cases, we can confirm that our feature construction scheme benefits the graph models' performance. The lower results of \textit{Zero} on traditional MPNNs tells us that these primitive architectures are misguided by the missing features; whereas the more modern graph transformers are robust to the absence of information, resulting in a better utilization of these partially-available features. These findings are further corroborated by a more comprehensive set of experiments below.

\noindent{\textbf{RQ2:} \textit{Are all features needed 
for improving model robustness?}}

Further delving into the contribution of each of our components, we conduct an ablation study on the feature configurations and reports the result in  \cref{tab:ablation}.
We select the commonly-used LDP features as our baseline, experiment with replacing or adding features, and measure the resulting differences. Note that \textit{Trim} does not work with LLM features: for all Android packages, there exist some functions that do not have any code in the APK (e.g. Android APIs). As a result, code embedding is not a universal feature, and thus will always be trimmed.

\vspace{-0.5ex}
\begin{finding}
For any feature and collation configurations, all models achieved higher results than the baseline.
\end{finding}
\vspace{-0.5ex}

This strengthens our claim of effectiveness for semantic features with these two collation schemes. 
As different GNNs response differently to each pipeline configuration, a practitioner may use these results to select an appropriate set of hyperparameters for their specific use cases.

\vspace{-0.5ex}
\begin{finding}
Transformer-based architectures achieve a higher in-distribution accuracy than MPNNs, but performs worse on distribution-shifted data.
\end{finding}
\vspace{-0.5ex}

During training, the models learn from the training distribution, and their final checkpoints are selected based on the corresponding validation set. This naturally results in these models fitting as much as possible to said distribution,
as the early-stopping only prevents overfitting to the training \textit{set}, not the distribution itself.
Given that the Transformer models have a higher discriminative power than MPNNs, they learned to work better on the trained task in exchange for a comparatively lower performance on covariate-shifted data.

\begin{table}[t]
\centering
\scriptsize
\caption{\small Accuracy on \textcolor{RoyalPurple}{MalNet-Tiny-Common} when combining TTA methods with semantic features. Highest values for each set in bold.}
\vspace{-0.5ex}
\renewcommand{\arraystretch}{0.7}
\begin{tabularx}{\columnwidth}{lYYYY}
\toprule
{Model} & {GCN} & {GIN} & {GPS} & {Exphormer} \\
\midrule
Tent & 44.90\% & 47.80\% & 47.90\% & 46.60\% \\
\textover[c]{\texttt{...}}{T3A} + Prune & \textbf{61.70\%} & \textbf{60.70\%} & 52.00\% & 52.20\% \\
\textover[c]{\texttt{...}}{T3A} + Zero & 58.50\% & 56.20\% & \textbf{53.80\%} & \textbf{55.70\%} \\
\midrule
T3A & 48.80\% & 46.80\% & 48.50\% & 47.90\% \\
\textover[c]{\texttt{...}}{T3A} + Prune & \textbf{63.40\%} & \textbf{62.00\%} & 52.00\% & 55.60\% \\
\textover[c]{\texttt{...}}{T3A} + Zero & 59.50\% & 57.70\% & \textbf{56.50\%} & \textbf{56.50\%} \\
\midrule
GTrans & 47.50\% & 47.20\% & 48.50\% & 50.10\% \\
\textover[c]{\texttt{...}}{T3A} + Prune & \textbf{62.00\%} & \textbf{61.40\%} & \textbf{55.00\%} & 56.50\% \\
\textover[c]{\texttt{...}}{T3A} + Zero & 58.90\% & 57.70\% & 54.40\% & \textbf{59.70\%} \\
\bottomrule
\end{tabularx}
\label{tab:dg_tta}
\vspace{-2.5ex}
\end{table}

\vspace{-0.5ex}
\begin{finding}
With adequate semantic context, simple MPNNs can achieve near the performance of Transformer-based architectures, while performing better under distribution shift.
\end{finding}
\vspace{-0.5ex}

For upstream classification, this phenomenon is best shown where GCN, a basic architecture, 
achieved 
9.5\% improvement in upstream task, outperforming the best Exphormer result. Similarly, GIN, a discriminative-oriented MPNN, increased its distribution-shifted accuracy by 14.2\%, far higher than the Transformer-based counterparts. These results show that the inherent lower representation capacity of primitive architectures can be more than compensated with good feature constructions for the input graph.
Additionally, we can conclude that our pipeline not only improves the models' general prediction capacity, but also 
their robustness. This is evident by the much larger gain in downstream accuracy comparing to which evaluated on the upstream distribution.

\begin{figure}[t]
\centering
\includegraphics[width=\linewidth]{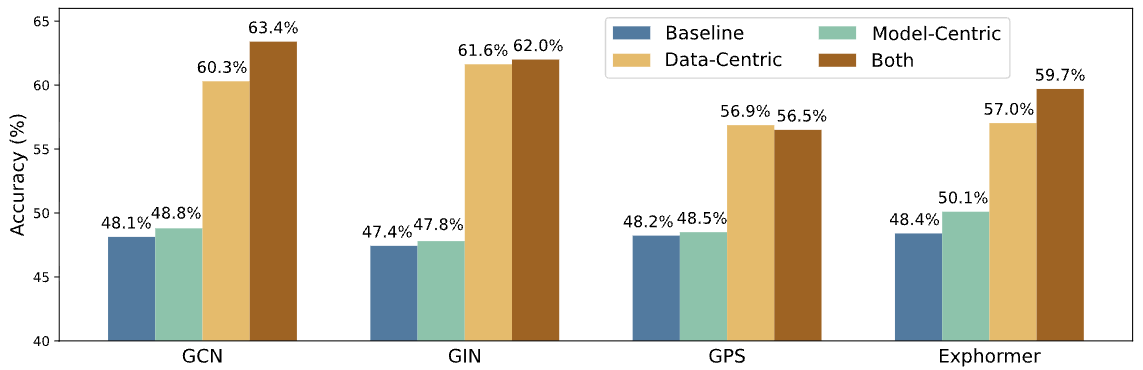}
\vspace{-3.9ex}
\caption{\small Models' accuracy under covariate shift after applying model-centric adaptation and/or data-centric augmentation.
}
\label{fig:model-vs-data}
\vspace{-1ex}
\end{figure}

\subsubsection{Adaptation-Based Training for Distribution Shift}
We plot the best accuracy achieved 
using existing model-centric approaches and our data-centric augmentation when evaluating on MalNet-Tiny-Common in~\cref{fig:model-vs-data}. Our feature construction method consistently beats TTA methods on their own, and in all cases but one architecture further improve OOD performance when used in conjunction with TTA methods.

For a more in-depth analysis of the complementary effect of these two paradigms, we evaluate the OOD performance of different GNNs when combined with each TTA method. \cref{tab:dg_tta} reports our results,
which show that most test-time approaches do not help generalizing to these new variants.

\noindent{\textbf{RQ3:} \textit{Do semantic features work with adaptation methods?}}

\vspace{-0.5ex}
\begin{finding}
Our method consistently improves the performance of the adapted models across different generic TTA methods and architectures. Specifically, \textit{Prune} works best with MPNNs while \textit{Zero} works best with Transformers.
\end{finding}
\vspace{-0.5ex}

Besides a higher accuracy across all experiments, we note that our previous findings on the effects of architectural differences are consistent in the TTA scenario. Specifically, \textit{Prune} continues to work best on MPNNs, delivering higher covariate-shifted results when compared to Transformers, which are more suitable to the \textit{Zero} collation scheme.

\begin{table}[t]
\small
\centering
\scriptsize 
\caption{\small Full dataset accuracy on MalNet-Tiny variants (\textcolor{RoyalPurple}{Common}, \textcolor{ForestGreen}{Distinct}) of models trained on MalNet-Tiny after finetuning.
}
\vspace{-0.5ex}
\renewcommand{\arraystretch}{0.75}
\begin{tabularx}{\columnwidth}{lYYYYYYYY}
\toprule
\multirow{2}{*}{Method} & \multicolumn{2}{c}{GCN} & \multicolumn{2}{c}{GIN} & \multicolumn{2}{c}{GPS} & \multicolumn{2}{c}{Exphormer} \\
\cmidrule(lr){2-3} \cmidrule(lr){4-5} \cmidrule(lr){6-7} \cmidrule(lr){8-9}
& \textcolor{RoyalPurple}{Cmn.} & \textcolor{ForestGreen}{Dst.} & \textcolor{RoyalPurple}{Cmn.} & \textcolor{ForestGreen}{Dst.} & \textcolor{RoyalPurple}{Cmn.} & \textcolor{ForestGreen}{Dst.} & \textcolor{RoyalPurple}{Cmn.} & \textcolor{ForestGreen}{Dst.} \\
\midrule
Finetune & 81.4\% & 94.8\% & 86.4\% & 96.3\% & 92.7\% & 97.2\% & 93.3\% & 97.8\% \\
\textover[c]{\texttt{...}}{T3A} + Prune & \textbf{96.8\%} & 96.8\% & \textbf{94.7\%} & 97.2\% & 96.9\% & \textbf{98.7\%} & \textbf{97.1\%} & 97.8\% \\
\textover[c]{\texttt{...}}{T3A} + Zero & 96.0\% & \textbf{97.2\%} & \textbf{94.7\%} & \textbf{97.3\%} & \textbf{97.0\%} & 98.4\% & 96.8\% & \textbf{97.9\%} \\
\midrule
k-NN Probe & 75.1\% & \textbf{90.3\%} & 75.0\% & \textbf{87.4\%} & 71.0\% & 87.5\% & 71.6\% & 87.5\% \\
\textover[c]{\texttt{...}}{T3A} + Prune & \textbf{85.7\%} & 86.3\% & 80.1\% & 85.7\% & 78.8\% & 89.2\% & 76.4\% & \textbf{90.4\%} \\
\textover[c]{\texttt{...}}{T3A} + Zero & 84.2\% & 89.8\% & \textbf{80.7\%} & 84.5\% & \textbf{78.9\%} & \textbf{90.5\%} & \textbf{79.5\%} & 89.6\% \\
\midrule
AdapterGNN & 80.2\% & 94.9\% & 83.8\% & \textbf{95.7\%} & 87.3\% & 95.6\% & 87.1\% & \textbf{96.9\%} \\
\textover[c]{\texttt{...}}{T3A} + Prune & \textbf{91.8\%} & \textbf{95.3\%} & \textbf{87.2\%} & 91.1\% & \textbf{93.7\%} & \textbf{96.6\%} & \textbf{92.8\%} & 96.2\% \\
\textover[c]{\texttt{...}}{T3A} + Zero & 90.7\% & 95.1\% & 86.9\% & 91.8\% & 92.4\% & 95.8\% & 91.2\% & 96.5\% \\

\bottomrule
\end{tabularx}
\vspace{-3ex}
\label{tab:dg_ft}
\end{table}

\noindent\textbf{Finetuning-based adaptation.}
To better improve performance under severe distribution shifts, some methods opt for adapting the model weights directly assuming the availability of additional labeled OOD data. While this requirement is unrealistic in mitigating zero-day attacks, it can be applicable to a post-hoc update to an existing detector. In this setting, all DA methods reach near the standard finetuning baseline (\cref{tab:dg_ft}).

\vspace{-0.5ex}
\begin{finding}
Our method consistently further improves the adapted models across different finetuning approaches.
\end{finding}
\vspace{-0.5ex}

The performance on the covariate-shifted dataset still consistently increase by up to 15.4\% on GCN, rivaling modern architectures in all cases. This observation confirms the intuition that simpler models that fit less closely to the training distribution can be more easily adapted to a new downstream task.
On the contrary, the only exception arises when evaluating on the domain-shifted dataset, where DA methods sometimes work better without our semantic features.
With low/no tunable parameters, these approaches become less capable the handle the more information we provide the model. In practice however, any distribution update is applied using low-epoch finetuning, in which our pipeline always provide superior results comparing to all other settings.
\section{Related Work}
\label{sec:related_works}
\textbf{Android Malware Graph.}
Most works on malware graphs also rely on FCGs for structure~\cite{malhotra2024comparison}.
Older works are based on random walks~\cite{amdroid}, while some other adapted NLP methods to feature extraction~\cite{cgdroid,feng2023bejagnn}. Other methods such as
\cite{8752028} extracted CFGs as the representations instead, and focused on the Internet-of-Thing domain. 
Meanwhile, \cite{https://doi.org/10.1155/2022/7245403} took a different approach and extracted the abstract syntax tree.

\textbf{Graph Distribution Shift.}
The effect of distribution shift in graphs can be attacked from many angles; some finetuning-based~\cite{gui2023gadapterstructureawareparameterefficienttransfer,li2023adaptergnnparameterefficientfinetuningimproves} while others used domain adaptation methods~\cite{chen2022graphttatesttimeadaptation,wang2022testtimetraininggraphneural,jin2023empowering,ju2023graphpatchermitigatingdegreebias,hsu2025structuralalignmentimprovesgraph}. A more novel approach to graph adaptation is graph prompt tuning~\cite{liu2023graphpromptunifyingpretrainingdownstream,fang2024universalprompttuninggraph,fu2025edgeprompttuninggraph}; however, these methods typically require pretraining on large corpora. We refer interested readers to~\cite{zhang2024survey} for further reading on these topics.
\section{Conclusion}\label{sec:conclusion}
In this work, we introduced two new datasets, MalNet-Tiny-Common and MalNet-Tiny-Distinct, to evaluate the robustness of Android malware classifiers against covariate and domain shifts, respectively.
Additionally, we constructed a semantic enrichment framework for Android function call graphs using function metadata and LLM code embeddings, demonstrating the effectiveness of our data-centric methodology in improving resilience of graph-based classifiers under distribution shift while maintaining strong performance on standard evaluation splits, especially in conjunction with existing model-based adaptation methods.
We hope this work will enable further research on enriching input graphs to improve the generalization of malware detectors and
stimulate future works on
semantics-aware graph representations for evolving threat environments.

\section*{Acknowledgement}
This material is supported by the National Science Foundation under Award Numbers 2325416, 2325417, and 2622415.

\clearpage
\bibliographystyle{IEEEtran}
\bibliography{ref}

\clearpage
\appendix
\subsection{Source Code and Dataset}
All code used to produce our results can be found at
https://github.com/ngoctnq/malnet-features.
The packaged PyG datasets, with instructions on how to use it, are available at
https://huggingface.co/datasets/ngoctnq/malnet-features.

\section{Dataset Construction Specifications}

\subsection{MalNet Function Metafeature Specifications}
\label{sec:metafeats}

We construct function node metafeatures by adapting the EMBER feature set~\cite{anderson2018emberopendatasettraining} to Android malwares, on a per-function basis. To check for storage access, we search for strings such as ``/storage/'' or ``/sdcard/''. For registry access equivalence, we look for ``/shared\_prefs/'', ``Settings.Secure'', ``Settings.System'', and ``Settings.Global''. MZ dropper is substituted with any sign of in-memory code execution, which exhibits in keywords such as ``ClassLoader'', ``DexFile'', ``loadDex'', ``loadClass'', “defineClass”, or ``loadLibrary''.

We also utilize method information available to us from Androguard~\cite{desnos2018androguard} analyzer getter functions. All numerical features are kept as is without normalization, and any string/string list features are converted into a 50-dimensional number vector using the hashing trick~\cite{weinberger2009feature}. We list all extracted features in Tab. \ref{tab:features}. Note that only the first 5 features in the table are available across all methods: for example, we cannot extract any bytecode statistics from external functions as they are not defined/available in the extracted APK. These 5 features form the \textbf{Trim} variant as described in the main text.

\subsection{Edge Features for Malware FCGs}
For edge features, we note that the relationship between two functions is an invocation, and thus any information about it (e.g. passed parameters, return values) requires an inspection of the call stack, which is only available at run time~\cite{dmalnet}. As we do not conduct dynamic analysis in this work, we do not extract any edge features to enrich the FCG representation.

\section{Detailed Experiment Setup}
\label{sec:supp_xp_details}

\subsection{Implementation Modifications}

\subsubsection{Readout function}
After message-passing, we use global max pooling to aggregate node features into a single graph-level representation, as also used in Transformer-based architectures~\cite{rampášek2023recipegeneralpowerfulscalable, shirzad2023exphormersparsetransformersgraphs}.
This readout choice has a nice interpretation: a program is the product of all its functions, and thus if one function behaves like a malware, the whole program is likely malicious. This is in contrast to the more traditional global mean pooling operation, which may dilute the effect of a single malicious function by averaging it with other benign functions.

\subsubsection{Classifier initialization} Across all experiments on MalNet-Tiny-Common, we start finetuning on the checkpoint as is. For MalNet-Tiny-Distinct, we reinitialize the classifier head for finetuning-based methods, as the model cannot adapt to new malware families without retraining the classifier.

\subsubsection{Prune} For \textit{Prune}, we omit removing isolated nodes during data preparation to prevent empty graphs. This is because for some malwares, all of its code-containing functions do not call each other (and e.g. only call APIs), and thus become completely isolated after all other nodes are pruned. 

\subsubsection{GTrans} We adapt the method as-is to graph classification by keeping all perturbation schemes and hyperparameters unchanged from the original code onto our graph classifiers, which only differ from their node classifying models in that the former has a readout step before classification. As some graph architectures did not support edge features, adjacency perturbation is disabled for a fair comparison.

\subsection{Hardware and Implementation}
All experiments are conducted on a single NVIDIA RTX 6000 Ada with 48Gb of memory. All runs are seeded with the same seed for reproducibility, using the hyperparameter configurations listed in the original paper~\cite{rampášek2023recipegeneralpowerfulscalable, shirzad2023exphormersparsetransformersgraphs}.
Each of our experiments takes 1-2 hours to run per seed, depending on the model architecture in use.

\section{Additional Experiment Results}
\label{sec:supp_add_xp}

\subsection{Pipeline Running Time}

\begin{table}[t]
\centering
\caption{\small Inference time of different architectures on MalNet-Tiny-Features comparing to original structure-only baseline.}
\begin{tabularx}{\linewidth}{cYYYY}
\toprule
Method & GCN & GIN & GPS & Exphormer \\
\midrule
Baseline & 
1x & 1x & 1x & 1x \\
\midrule
Trim  & 1.31x & 1.47x & 1.02x & 1.12x \\
Prune & 2.65x & 3.69x & 1.12x & 1.24x \\
Zero  & 4.51x & 6.20x & 1.25x & 2.34x \\
\bottomrule
\end{tabularx}
\label{tab:runtime}
\vskip -1em
\end{table}

\Cref{tab:runtime} reports the comparative evaluation time across different variants of our approach. The multipliers are as expected: GCN layer only include a matrix multiplication, which scales linearly with the number of features; while GIN scales quadraticly due to having an MLP at every layer. The Transformer architectures, in contrast, have a feature downscaler very early on before the main computational-heavy components, and thus the difference in overhead with respect to dimension size is much more negligible.

This graceful scaling despite the large feature set (upto 651x vs. baseline for our largest variant) is thanks to the approaches of our method in dealing with non-universal features, which take up a large portion of the feature matrix. Specifically, \textit{Prune} removes up to 89.60\% of nodes that do not have all available features (even when it keeps full-featured isolated nodes while the other datasets have them removed); and \textit{Zero} creates a sparse matrix with up to 89.09\% of zeros per node. These properties drastically reduce the amount of computation needed for such a high-dimensional input to the models.

\begin{table}[t]
\small 
\centering
\caption{\small Specifications of all extracted function metafeatures.}
\centering
\begin{tabularx}{\linewidth}{p{1.3cm}Xll}
\toprule
\multicolumn{2}{c}{Feature} & Type & Embedding \\
\midrule
\multirow{2}{20mm}{Names} & Class name & [String]\tnote{[a]} & Hashing trick \\
& {Method name} & string & Hashing trick \\
\midrule
\multirow{3}{40mm}{Method\\signature} & Number of parameters & Integer & As-is \\
 & Parameter types & [String] & Hashing trick \\
 & Return type & string & Hashing trick \\
\midrule
\multirow{2}{20mm}{Method\\misc.} & Access flags & Binary\tnote{[b]} & Multi-hot \\
 & No. of local registers & Integer & As-is \\
\midrule
\multirow{3}{20mm}{Code} & Length & Integer & As-is \\
 & Byte histogram & [Integer] & Distribution\tnote{[c]} \\
 & Byte-entropy hist. & [Integer] & Distribution \\
\midrule
\multirow{2}{0mm}{Instructions} & Length & Integer & As-is \\
 & Opcode names & String list & Hashing trick \\
\midrule
\multirow{11}{20mm}{Strings} & Contains invalid char. & Boolean & As-is \\
 & String literal & String & Hashing trick \\
 & Number of strings & Integer & As-is \\
 & Average string length & Float & As-is \\
 & Character histogram & [Integer] & Distribution \\
 & Character entropy & Float & As-is \\
 & No. of external paths & Integer & As-is \\
 & Number of URLs & Integer & As-is \\
 & No. of IP addresses & Integer & As-is \\
 & Registry modification & Integer & As-is \\
 & In-memory executions & Integer & As-is \\
\midrule
Misc. & Instructions cached? & Boolean & As-is \\
\bottomrule \\
\end{tabularx}
\vskip -1em
\label{tab:features}
\end{table}

\subsection{Different LLM Options}
\label{sec:llm_ablation}
\begin{table}[t]
\scriptsize
\centering
\caption{\small Models' accuracy on \textcolor{RoyalPurple}{MalNet-Tiny-Common} across different LLM emedding models. Subscript denotes standard deviation over 3 independent runs. Best result in bold, second best underlined.}
\setlength{\tabcolsep}{4pt}
\begin{tabularx}{\linewidth}{c!{\color{LightGray}\vrule width 0.1pt}ccc!{\color{LightGray}\vrule width 0.1pt}YYYY}
\toprule
& Meta & LLM & LDP & GCN & GIN & GPS & Exphormer \\

\midrule
\multirow{14}{*}{\rotatebox{90}{Prune}} & \greyno & CXE & \greyno & $\mathbf{60.30}_{1.98}$ & $\mathbf{61.10}_{0.22}$ & $53.50_{2.12}$ & $\mathbf{55.27}_{0.66}$ \\
& \greyno & UniX & \greyno & $54.90_{0.85}$ & $58.77_{1.94}$ & $\mathbf{56.43}_{1.08}$ & \underline{$54.93_{0.50}$} \\
& \greyno & Qwen & \greyno & \underline{$55.23_{2.08}$} & \underline{$59.97_{0.54}$} & \underline{$54.57_{1.04}$} & $53.27_{1.20}$ \\
\cmidrule(lr){2-8}
& \greyno & CXE & \checkmark & $\mathbf{58.33}_{1.72}$ & $\mathbf{59.97}_{0.95}$ & $53.07_{0.73}$ & \underline{$54.87_{1.52}$} \\
& \greyno & UniX & \checkmark & $55.23_{0.69}$ & $56.63_{0.69}$ & $\mathbf{56.53}_{0.56}$ & $\mathbf{55.27}_{0.95}$ \\
& \greyno & Qwen & \checkmark & \underline{$57.57_{1.38}$} & \underline{$58.43_{1.08}$} & \underline{$54.07_{0.95}$} & $52.57_{1.45}$ \\
\cmidrule(lr){2-8}
& \checkmark & CXE & \greyno & $\mathbf{59.70}_{0.41}$ & $\mathbf{61.63}_{2.19}$ & $\mathbf{55.57}_{1.52}$ & \underline{$52.37_{0.12}$} \\
& \checkmark & UniX & \greyno & $53.73_{1.09}$ & $54.97_{0.73}$ & $54.30_{0.22}$ & $\mathbf{52.93}_{0.83}$ \\
& \checkmark & Qwen & \greyno & \underline{$55.60_{1.07}$} & \underline{$59.13_{0.25}$} & \underline{$54.73_{0.98}$} & $52.00_{0.78}$ \\
\cmidrule(lr){2-8}
& \checkmark & CXE & \checkmark & $\mathbf{57.03}_{1.08}$ & $\mathbf{60.07}_{1.67}$ & $\mathbf{56.70}_{0.93}$ & $\mathbf{53.93}_{1.05}$ \\
& \checkmark & UniX & \checkmark & $55.73_{1.41}$ & $55.33_{0.45}$ & $52.50_{1.20}$ & \underline{$53.23_{1.47}$} \\
& \checkmark & Qwen & \checkmark & \underline{$57.00_{2.00}$} & \underline{$57.40_{0.36}$} & \underline{$55.07_{1.68}$} & $51.87_{1.35}$ \\
\midrule
\multirow{14}{*}{\rotatebox{90}{Zero}} & \greyno & CXE & \greyno & \underline{$56.70_{2.03}$} & $\mathbf{58.60}_{1.70}$ & $\mathbf{56.73}_{0.54}$ & $55.57_{1.58}$ \\
& \greyno & UniX & \greyno & $56.17_{1.43}$ & \underline{$58.07_{1.52}$} & \underline{$55.17_{0.09}$} & $\mathbf{56.07}_{1.06}$ \\
& \greyno & Qwen & \greyno & $\mathbf{57.93}_{0.82}$ & $57.93_{2.25}$ & $54.07_{2.00}$ & \underline{$55.70_{2.09}$} \\
\cmidrule(lr){2-8}
& \greyno & CXE & \checkmark & $\mathbf{58.60}_{1.19}$ & $56.57_{0.34}$ & $\mathbf{54.87}_{1.39}$ & $\mathbf{55.33}_{2.33}$ \\
& \greyno & UniX & \checkmark & $53.30_{0.50}$ & $\mathbf{57.60}_{1.19}$ & $53.87_{1.86}$ & \underline{$54.27_{0.74}$} \\
& \greyno & Qwen & \checkmark & \underline{$56.13_{0.91}$} & \underline{$57.33_{0.21}$} & \underline{$53.93_{2.04}$} & $51.77_{1.51}$ \\
\cmidrule(lr){2-8}
& \checkmark & CXE & \greyno & $\mathbf{58.87}_{1.46}$ & \underline{$56.33_{0.62}$} & \underline{$54.97_{1.48}$} & $\mathbf{55.87}_{0.31}$ \\
& \checkmark & UniX & \greyno & $52.53_{1.19}$ & $56.13_{0.33}$ & $52.00_{1.49}$ & \underline{$55.83_{1.43}$} \\
& \checkmark & Qwen & \greyno & \underline{$53.97_{1.21}$} & $\mathbf{59.23}_{0.66}$ & $\mathbf{55.77}_{1.62}$ & $54.03_{0.45}$ \\
\cmidrule(lr){2-8}
& \checkmark & CXE & \checkmark & $\mathbf{59.00}_{0.86}$ & $\mathbf{59.10}_{0.67}$ & \underline{$53.60_{0.67}$} & $54.73_{1.43}$ \\
& \checkmark & UniX & \checkmark & \underline{$53.37_{1.09}$} & $56.60_{1.12}$ & $\mathbf{54.90}_{1.02}$ & \underline{$54.90_{0.64}$} \\
& \checkmark & Qwen & \checkmark & $53.23_{0.45}$ & \underline{$57.13_{0.78}$} & $53.30_{2.12}$ & $\mathbf{55.70}_{1.53}$ \\

\bottomrule
\end{tabularx}
\vskip -1em
\label{tab:llm_ablation}
\end{table}

In this section, we conduct our evaluations with different LLMs for feature extractions. Besides CodeXEmbed~\cite{liu2024codexembedgeneralistembeddingmodel} which is the only LLM that satisfies all of our requirements, we also experiment with UniXcoder~\cite{guo2022unixcoderunifiedcrossmodalpretraining} (UniX), which is a code-specific LLM embedder with a much smaller context window of 512; and Qwen 3~\cite{yang2025qwen3technicalreport}, a multi-domain LLM which can support data in code domain with a larger context.

The complete ablation results are reports in \Cref{tab:llm_ablation}, in which CodeXEmbed (CXE) outperforms the other LLMs across most variants of the pipeline as we expected. We thus opt for CodeXEmbed as the default option for our method.

\section{Discussions}

\subsection{Practical Applicability}
\label{sec:practical}
The empirical results presented in \cref{sec:results} provide a clear decision-making framework for deploying robust malware detection systems. We concisely identify three key dimensions for practical implementation.

\subsubsection{Model Selection: Speed vs. Utility vs. Robustness}
As different configurations behave differently under the same data distribution, with the results from our ablation studies, one can make an informed decision considering the tradeoffs for their specific needs. We provide here some general suggestions:
\begin{itemize}[leftmargin=*]
    \item For resource constraints, we recommend using GIN-\textit{Prune} with only metafeatures. GIN is the smallest and fastest architecture, and we have negligible computation overhead: the time needed to process metafeatures is heavily dominated by the time needed to deconstruct the APK, which is required during FCG extraction. If LLM features are desired for a more balanced tradeoff, smaller LLMs can be used.
    \item For peak utility, GPS-\textit{Zero} with only semantic features (i.e. no LDP) yield the highest in-distribution test accuracy. Note that GPS is very slow, roughly 10.3 times slower than GIN.
    \item For maximum robustness, GIN-\textit{Prune} with only semantic features performed the best under covariate shift, trading off only 1\% upstream accuracy for a 6.67\% gain under OOD.
\end{itemize}
\subsubsection{Continuous Adaptation to New Threats}
Continuous updates have been the proper practice against the rapid evolvement of malwares, with antiviruses frequently updating their definition databases. Equivalently for the ML-powered counterpart, the classifier's weights should also be frequently updated whenever new malware variants are discovered, given that domain adaptation improved the most for out-of-distribution performance (c.f. \cref{tab:dg_ft}). While DA approaches can be costly in terms of computation, it can be justified by amortizing by the large number of benefiting end users. Additionally, in this scenario, Exphormer should be the preferred choice of architecture if accuracy is of utmost importance.
\subsubsection{Handling Low-Quality Data Samples}
In real-world deployment, malware binaries are often obfuscated or rely heavily on external system libraries that lack source code for LLM embedding. Our framework's ability to handle partially-defined graphs (\cref{sec:feature_collation}) ensures that the system does not fail when features are missing. A final note, in the extreme case where the malware is fully composed of externally-compiled functions (and thus decompilation is impossible), we suggest opting for metafeature-only configuration to not dilute the model's attention to these unavailable features.

\subsection{Ethics Statement}
The data source used in this work are readily-available applications on AndroZoo~\cite{androzoo}, collected and derived with explicit permission of the original maintainer. All samples were pre-processed to remove any Personally Identifiable Information and raw executable code~\cite{anderson2018emberopendatasettraining}. Our release contains only extracted features representing the software's control flow and the general semantic properties of the functions, ensuring that no sensitive user data or proprietary source code is exposed. The representations are benign and cannot be executed to perform malicious actions. Labeling is provided by MalNet~\cite{freitas2021largescaledatabasegraphrepresentation} with a CC-BY license, permitting modification to the data with proper attribution.

\end{document}